\documentclass[superscriptaddress,twocolumn,showpacs,prl,floatfix]{revtex4}

\usepackage{color}
\usepackage{tabularx}
\usepackage{epsfig}
\usepackage{amsmath}
\usepackage{graphicx}

\begin{document}

\title{Dynamics in the Sherrington-Kirkpatrick Ising spin glass at and above $T_{g}$}

\author{Alain Billoire}
\affiliation {Institut de physique th\'eorique, CEA Saclay and CNRS, 91191 Gif-sur-Yvette, France}

\author{I. A.~Campbell}
\affiliation{Laboratoire Charles Coulomb,
  Universit\'e Montpellier II, 34095 Montpellier, France}

\begin{abstract}
A detailed numerical study is made of relaxation at equilibrium in the Sherrington-Kirkpatrick Ising spin glass model, at and above the critical temperature $T_{g}$. The data show a long time stretched exponential relaxation
$q(t) \sim \exp[-(t/\tau(T))^{\beta(T)}]$ with an exponent $\beta(T)$ tending to $\approx 1/3$ at $T_{g}$. The results are compared to those which were observed by Ogielski in the $3d$ ISG model, and are discussed in terms of a phase space percolation transition scenario.
\end{abstract}


\maketitle

\section{Introduction}

The Sherrington-Kirkpatrick (SK) Ising spin glass (ISG) model has been intensively studied for almost forty years. Introduced \cite{sherrington:75,kirkpatrick:78} as a starting point for studying Edwards-Anderson-like \cite{edwards:75} spin glasses, it is a classical mean-field $N$-spin Ising model in which there are quenched random interactions between all pairs of spins.
It has an ordered spin glass phase below the critical temperature $T_{g}=1$. The static properties such as the specific heat, the magnetic susceptibility and the spin glass order parameter are known to high precision in the ordered state and are well understood in terms of the Parisi Replica Symmetry Breaking (RSB) theory \cite{parisi:79} and its subsequent developments. The paramagnetic regime has been considered "trivial" in that there are simple exact expressions for the basic static physical properties in the large $N$ limit.

In the ordered state the dynamics are very slow; the equilibrium relaxation time diverges exponentially  when the number of sites $N$ goes to infinity \cite{billoire:01,billoire:10,billoire:11}. For the paramagnetic state, the dynamics were discussed in early work,
Refs. \cite{kinzel:77,kirkpatrick:78,sompolinski:81,sompolinski:82}. The ISG relaxation corresponds to a continuous spectrum of relaxation frequencies, for which an explicit linearized expression was given in Ref.~\cite{kirkpatrick:78}, leading to a relaxation function for Gaussian interactions and Glauber dynamics in the form of an integral, Ref.~\cite{kirkpatrick:78}, Eq. 5.27. It was concluded that in the large size limit at $T_{g}$ the relaxation at thermal equilibrium of the autocorrelation $q(t) =(1/N)\sum_{i} <S_{i}(0).S_{i}(t)>$ as a function of time $t$  (in Monte Carlo steps per spin) would be $q(t) \sim t^{-1/2}$, and above $T_{g}$ the very long time limit would take the form of a simple exponential $q(t) \sim \exp[-(t/\tau(T))]$ corresponding to a cutoff in the relaxation frequency spectrum.
As far as we are aware, after exploratory numerical calculations on a very limited number of Gaussian interaction samples over a few Glauber time steps in Ref.~\cite{kirkpatrick:78} no further numerical work on the dynamics in this temperature region has been reported.

Here a detailed numerical study of the thermal equilibrium relaxation of the autocorrelation function $q(t,N)$ in the ordered state as a function of time $t$ in updates per spin and $N$ the number of spins  \cite{billoire:11} is extended to temperatures at and above $T_{g}$. At $T_{g}$ the relaxation is always finite size limited; the expected scaling form
\begin{equation}
q(t,N)N^{1/3} = F[-(t/N^{2/3})]
\label{qtN33}
\end{equation}
is observed. For the three temperatures above $T_{g}$ at which measurements were made the large size limit behavior [$q(t,N)$ independent of $N$] is reached with the largest samples studied.

For $q(t)$ values down to at least $q(t) \approx 10^{-6}$ the autocorrelation function decay is strongly non-exponential. Satisfactory long $t$ fits are given both above $T_g$ and in the finite size scaling regime at $T_g$ by stretched exponentials
\begin{equation}
q(t) \sim exp[-(t/\tau(T))^{\beta(T)}]
\label{qt}
\end{equation}
with a temperature dependent exponent $\beta(T)$ which is smaller than $1$ and which tends to $\beta(T) \sim 1/3$ at $T=T_g$ \cite{beta}.

\section{The Mean field ferromagnet}

Before discussing the SK data analysis it is instructive as an illustration  to recall the (non-equilibrium) relaxation behavior for the mean field Ising ferromagnet, for which exact results by Suzuki and Kubo \cite{suzuki:68} were discussed in Ref. \cite{kirkpatrick:78}. At temperatures above $T_c=1$ the relaxation takes the form
\begin{multline}
m(t,T)=\\(1-\beta)^{1/2}/[(1-\beta +\beta^{3}/3)\exp(2(1-\beta)t)-\beta^3/3]^{1/2}
\label{ferro}
\end{multline}
where $\beta=1/T$, which becomes
\begin{equation}
m(t,T_c) = (1+ (2/3)t)^{-1/2}
\label{ferroTc}
\end{equation}
at $T_c$.
It is easy to see from Eq.~\ref{ferro} that above $T_{c}$ and beyond a time $t \simeq \tau(T) = 1/(1-\beta)$, $m(t,T)$ is essentially equal to $A(T)\exp(-t/\tau(T))$, a pure exponential with a time independent prefactor,
\begin{equation}
A(T) = [(1-\beta)/(1-\beta +\beta^{3}/3)]^{1/2}
\label{ferroA}
\end{equation}
see Figure 1.
Beyond $t \approx \tau$ only one single mode contributes to the relaxation in the mean field ferromagnet.
At short times (time zero to time $\sim \tau(T)$) there is a more complex function which connects up between the starting point $m(0,T)=1$ and the asymptotic region.
At no time for temperatures above $T_c$ is $m(t,T) \approx t^{-1/2}\exp(-t/\tau)$ a useful approximation.

It can be helpful to show the data in the form of a derivative plot $-d\log(m(t))/d\log(t)$ against $t$. Then a pure exponential $m(t)=A\exp[-t/\tau]$ appears as a straight line of slope $1/\tau$ passing through the origin.

\begin{figure}
\includegraphics[width=3.5in]{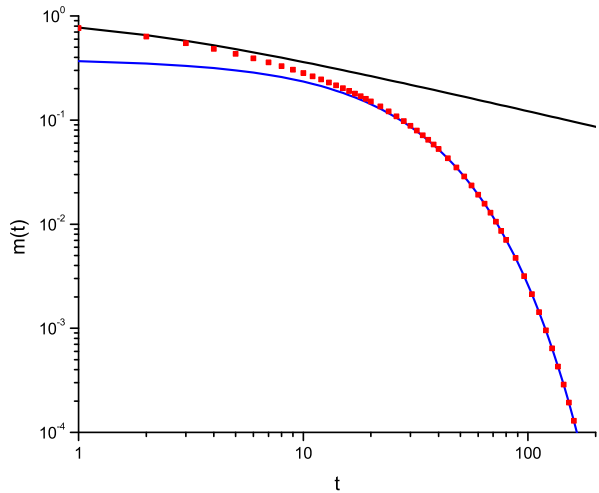}
  \caption{(Color online) An example of an $m(t)$ plot for the mean field ferromagnet calculated using Eq.~\ref{ferro}. The points correspond to $m(t)$ at $T=1.05T_{c}$; the black line is the algebraic critical behavior, and the blue curve is the asymptotic pure exponential.} \protect\label{fig:1}
\end{figure}

\section{SK simulations}

We now return to the SK simulations.
Systems of size $N = 64, 128, 256, 512, 1024$ and $2048$ were equilibrated using the procedure described in Ref. \cite{billoire:11}. Simulations were
made on $1024$ independent disorder samples at each size
with  two clones simulated in parallel. (As shown in Ref. \cite{billoire:10} two clones allow one to build up a noisy but unbiased estimate of the thermal fluctuations. Here they are only used to decrease the thermal fluctuations). We performed
$10^7$ Monte Carlo updates per site after thermalization, with measurements of $q(t)$ made every 4 time intervals. Both Metropolis and heat bath relaxation runs were carried out. It can be noted that SK simulations are numerically demanding : there are about as many individual spin-spin interactions in an $N=2048$ SK sample as in an $L=90$ sample in dimension $3$.

In the $N=256$ data set, and that set only, it turns out that one specific disorder sample $i$ gives $q_{i}(t,T)$ that are very far from the values for all the $1023$ other samples of the same size. This is true for sample $i$ with either Heat Bath or Metropolis updating schemes, and for all values of $T$. For example at $T=T_g=1$ we have $1024$ estimates for $q(256,t=1024)$ with a median equal to 0.0000040, first and third quartiles equal to -0.0001183 and 0.0001292 respectively, and an isolated maximum value equal to 0.2946040. This single outlier was omitted from the global analysis, since it would have led to a long $t$ distortion of our values of the mean $<q(256,t,T)>$ and the statistical error. At long $t$ when $<q(t,T)>$ is very small, if one single sample by statistical accident is very uncharacteristic it can have a big influence on the mean and the error. Note that in our simulation the thermal noise on $<q(t,T)>$ is much smaller than the disorder noise.

The temperatures at which measurements were made are $T = T_{g}= 1$ and $T = 1.1, 1.2, 1.3$

\section{SK analysis}

The standard dynamic finite size scaling rule exactly at the critical point can be written
\begin{equation}
q_{c}(L,t)= L^{-\beta/\nu} F[t/\tau(L)]
\end{equation}
with $\tau(L) \sim L^z$.
It is well known that quite generally finite size scaling is modified above the upper critical dimension $d_{u}$, with the combination $(T-T_g)L^{1/\nu}$ replaced by $(T-T_g)N^{1/(d_{u}\nu)}$ \cite{brezin}, where $N$ is the number of sites, namely $L$ is replaced by $N^{1/d_{u}}$.

The SK model is the infinite dimensional version of a model with upper critical dimension $d_{u}=6$ \cite{cirano}, and its scaling behavior is accordingly described by the effective length scale $L_{eff}=N^{1/6}$. This result can be obtained directly by noting that in the critical region the Landau expansion of the free energy of the replicated model starts (symbolically) by $a(T-T_g)Q^{2}+bQ^{3}$ with $Q$ the $n$ by $n$ matrix $q_{a,b}$ where $n$ is the number of replica. The partition function takes the form $\int dQ \exp(-N(a(T-T_g)Q^{2}+b Q^{3}))$, and after the change of variable $Q=X/[(T-T_g)N]^{1/2}$ one obtains a free energy that is a function of $(T-T_g)N^{1/3}$, in agreement with the above statement since $\nu=1/2$ for this model.

Therefore for dimension $6$ and
above (including SK) the effective ISG "length" scale is $L_{eff} = N^{1/6}$. The mean field ISG critical exponents are $\beta=1$, $\nu=1/2$, and $z=4$.

Hence in the SK case the dynamic scaling becomes
\begin{equation}
q_{c}(N,t) \sim N^{-1/3}F[t/N^{2/3}]
\label{qctN}
\end{equation}
so that in a critical temperature scaling plot one should expect $q_{c}(N,t)N^{1/3}$ to be a function $F[t/N^{2/3}]$.

In conditions where $q_{c}(N,t)$ approaches the $N \to \infty$ limit, the scaling rule becomes
\begin{equation}
q_{c}(t)\sim N^{-1/3}(t/N^{2/3})^{-1/2} = t^{-1/2}
\label{qct}
\end{equation}
independent of $N$. This is the situation for short times and large $N$, Figure 2.

\begin{figure}
\includegraphics[width=3.5in]{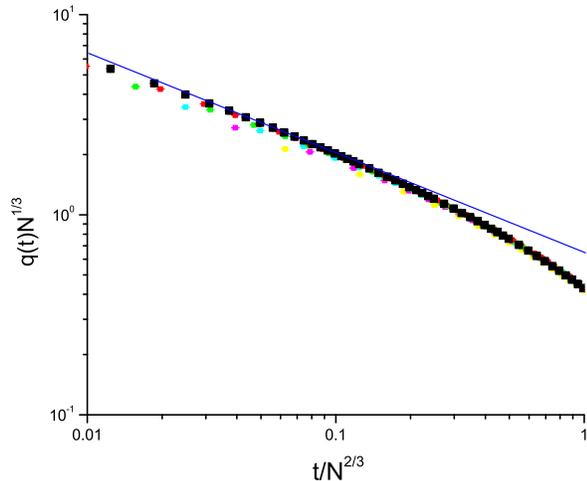}
  \caption{(Color online) The short time scaled SK relaxation data $q(t)N^{1/3}$ against $t/N^{2/3}$ in logarithmic coordinates at the critical temperature $T=1$. The colors for $N=64, 128, 256, 512, 1024, 2048$ are yellow, pink, cyan, green, red and black. For clarity the $N=2048$ points are larger than the others. The blue line indicates $q_c(t)\sim t^{-1/2}$.} \protect\label{fig:2}
\end{figure}

\begin{figure}
\includegraphics[width=3.5in]{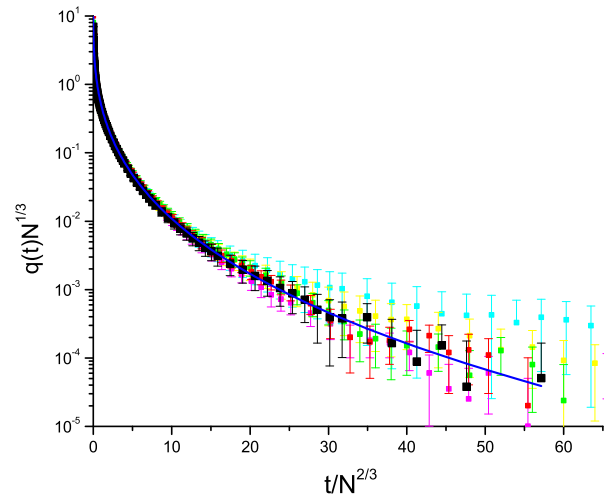}
  \caption{(Color online) The scaled SK relaxation data $q(t)N^{1/3}$ in logarithmic coordinates against $t/N^{2/3}$ at the critical temperature $T=1$. The color code is the same as in Figure 2. The blue curve is a stretched exponential $q(t)N^{1/3}=14\exp[-((t/N^{2/3})^{1/3})]$.} \protect\label{fig:3}
\end{figure}

For reasons to be discussed below, in addition to testing this scaling form we want to test if the long time critical finite size scaling function $F[x]$ is consistent with the stretched exponential form $B\exp[-(x)^{\beta}]$ having the particular exponent value $\beta =1/3$.
In figure 3 the scaled critical temperature relaxation data are plotted as $\log(q(t)N^{1/3})$ on the y-axis against $(t/N^{2/3})$ on the x-axis.
The long time critical scaling rule is well obeyed over the whole range of sizes $N$ which have been studied; beyond times of a few Monte Carlo steps $q_{c}(N,t)N^{1/3}$ is indeed an $N$-independent  function $F[t/N^{2/3}]$ to within the error bars for all $t$ and $N$, with no visible sign of corrections to finite size scaling.

The initial very short time relaxation (Figure 2, invisible in Figure 3) follows approximately the $N$-independent form $q(t) \sim t^{-1/2}$ as predicted.
As can be seen in Figure 3, within the numerical precision the finite size scaling function is from then on compatible with a stretched exponential having exponent $\beta = 1/3$.

For the three temperatures above $T_{g}$, $T = 1.1, 1.2, 1.3$
the raw $q(N,t)$ data are shown in Figures 4, 5 and 6.
It can be seen that at each of these temperatures the size dependence saturates with increasing $N$. The $q(N,t)$ curves for the largest sizes coincide within the numerical error bars, so these data can be taken to represent the "infinite" $N$ limit behavior.

The $q(t)$ curves calculated with the expression given in Ref.~\cite{kirkpatrick:78} Eq. 5.27, are almost identical to the limiting numerical $q(N,t)$ curves for $T=1.1$ and $T=1.2$ but the calculated curve for $T=1.3$ lies somewhat above the numerical data and has a rather different shape. The good quantitative agreement for the first two temperatures is perhaps fortuitous as the expression in Ref.~\cite{kirkpatrick:78} corresponds to a model with Gaussian interactions, and a linearization approximation was made in the calculation. The authors expected the expression to become accurate only at temperatures well above $T_g$. Nevertheless the approach of Ref.~\cite{kirkpatrick:78} is substantially validated by the numerical data.

We can note that the simulations made with the Metropolis updating rules (not shown) provide $q(t,N)$ data which are very similar to the heatbath data but with a global shift in the time scale to times shorter by a factor of between $2$ and $3$ depending on the temperature.
It can be shown that as a general rule $q(t)$ will fall faster with Metropolis updating than with heatbath/Glauber updating; the Metropolis acceptance rate is higher than that for heatbath/Glauber with a ratio of which depends on temperature.

\begin{figure}
\includegraphics[width=3.5in]{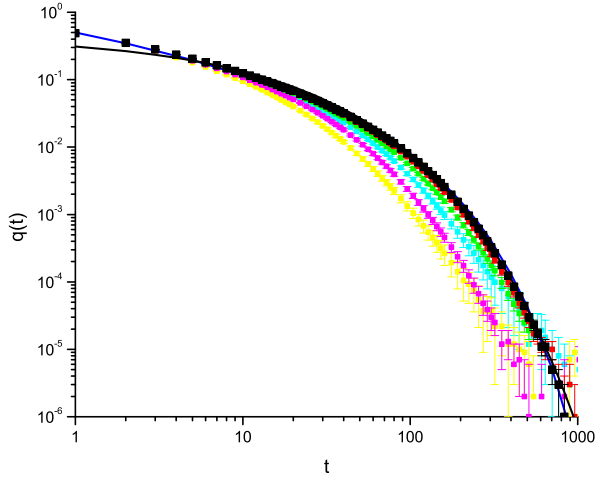}
  \caption{(Color online) The raw relaxation data $q(t)$ against $t$ at the temperature $T=1.1$. The color code for $N$ is the same as in Figure 2. The fit Eq.~\ref{qt} to the largest sizes is shown as the full black curve. $q(t)$ calculated from  Ref.~\cite{kirkpatrick:78} Eq. 5.27 is shown as the full blue curve.} \protect\label{fig:4}
\end{figure}
\begin{figure}
\includegraphics[width=3.5in]{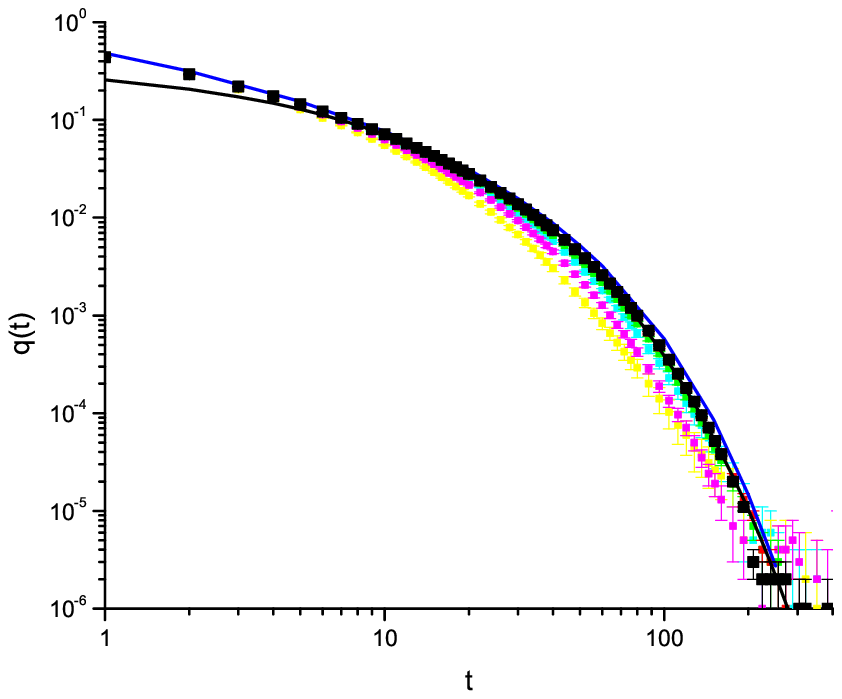}
  \caption{(Color online) The raw relaxation data $q(t)$ against $t$ at the temperature $T=1.2$. The color code for $N$ is the same as in Figure 2. The fit Eq.~\ref{qt} to the largest sizes is shown as the full black curve. $q(t)$ calculated from  Ref.~\cite{kirkpatrick:78} Eq. 5.27 is shown as the full blue curve.} \protect\label{fig:5}
\end{figure}
\begin{figure}
\includegraphics[width=3.5in]{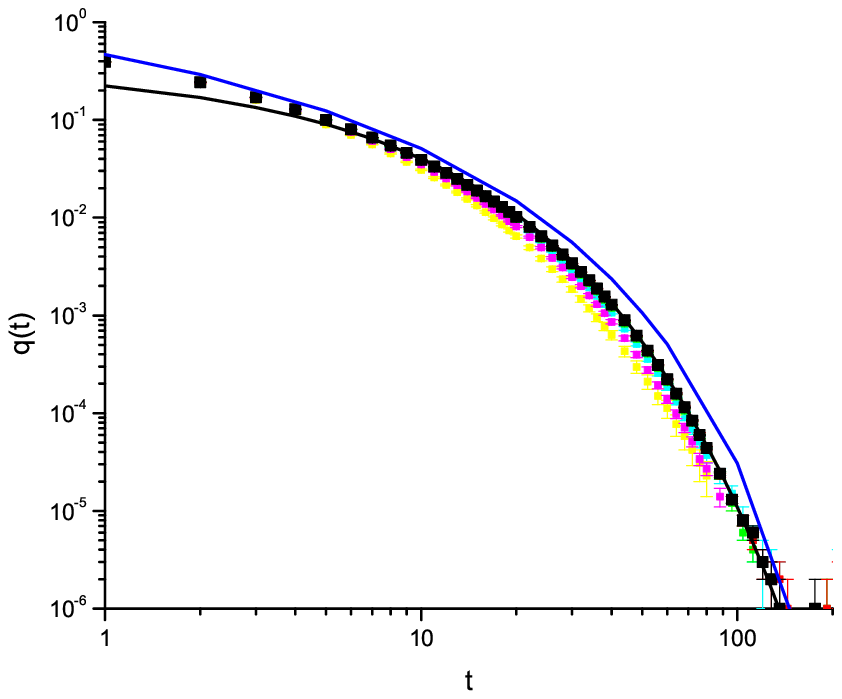}
  \caption{(Color online) The raw relaxation data $q(t)$ against $t$ at the temperature $T=1.3$. The color code for $N$ is the same as in Figure 2. The fit Eq.~\ref{qt} to the largest sizes is shown as the full black curve. $q(t)$ calculated from  Ref.~\cite{kirkpatrick:78} Eq. 5.27 is shown as the full blue curve.} \protect\label{fig:6}
\end{figure}

\begin{figure}
\includegraphics[width=3.5in]{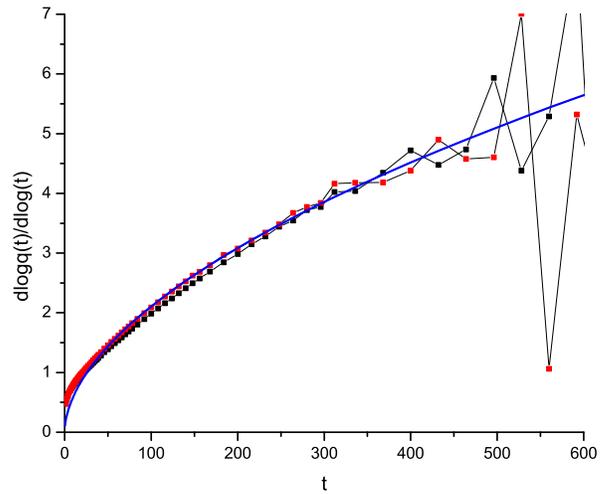}
  \caption{(Color online) The derivative plot $-d\log(q(t))/d\log(t)$ against $t$ at $T=1.1$ for the two largest sizes, $N = 2048$ (black) and $1024$ (red). The fit (blue curve) is for the same parameters as in Figure 4.} \protect\label{fig:7}
\end{figure}

The figures are of the same form as in the ferromagnetic example of Figure 1 with an initial short time regime followed by a long time asymptote, except that the long time ferromagnet pure exponential asymptote is replaced by a stretched exponential. Beyond the initial short time regime the stretched exponentials fit the large $N$ SK data to within the numerical precision.

If the stretched exponential form Eq.~\ref{qt} is assumed, then
\begin{equation}
-d\log(q(t))/d\log(t) =  \beta(T)(t/\tau)^{\beta(T)}
\label{dlogqt}
\end{equation}
An example of a plot of  $-d\log(q(t))/d\log(t)$ against $t$ is shown in Figure 7. For a pure exponential relaxation $\beta = 1$ the data points would lie on a straight line going through the origin. This is clearly not the case. By analogy with the ferromagnetic behavior, beyond an initial short time region the large $N$ data are fitted by a $B(T)t^{\beta(T)}$ curves with $\beta(T)$ less than $1$ and $B(T) = \beta(T)/\tau(T)^{\beta(T)}$.

Satisfactory fits are obtained with : $\beta(1.1) \approx 0.52$, $\tau(1.1) \approx 6.8$, $\beta(1.2) \approx 0.61$,  $\tau(1.2) \approx 4.2$,and $\beta(1.3) \approx 0.68$, $\tau(1.3) \approx 3.2$ for the heatbath updates. These fits are shown as the black curves in Figures 4 to 6.

The relaxation time $\tau(T)$ is of course increasing as $T_{g}$ is approached but with only three temperature points it is difficult to
estimate a functional form for $\tau(T)$. However the most significant result is that the relaxation above $T_{g}$ in the effectively infinite size limit can be represented by stretched exponentials having exponents $\beta(T)$ which are less than $1$ and which drop regularly as $T_{g}$ is approached.

From this result for the large $N$ limit relaxation data at the temperatures above $T_g$ together with the scaled finite size limited relaxation at criticality
it can thus be concluded that in the paramagnetic region the SK model thermodynamic limit equilibrium relaxation can be treated as essentially stretched exponential, with an exponent $\beta(T)$ which decreases continuously as the temperature is lowered. $\beta(T)$
tends to a value  $\beta_{c} \approx 1/3$ at criticality. It should be underlined that this behavior has been followed down numerically though five orders of magnitude in $q(t)$ at each temperature.

\section{ISG in dimension three}

It is instructive to compare with the situation for a finite dimension model. Ogielski \cite{ogielski:85} fitted thermodynamic limit $q(t)$ data on the bimodal ISG in dimension $3$ using an empirical function of the form
\begin{equation}
q(t) \sim t^{-x(T)}\exp[-(t/\tau(T))^{\beta(T)}]
\label{qtOg}
\end{equation}
where an algebraic prefactor multiplies a stretched exponential. Ogielski estimated $T_{g} \approx 1.175$ which is close to recent estimates $T_{g} \approx 1.12$ \cite{katzgraber:06,hasenbusch:08}. The values estimated for  the stretched exponential exponent $\beta(T)$ decrease regularly from $\beta(T) \approx 1$ at a temperature of about $4.0$ to $\beta(T) \approx 0.35$ at a temperature just above $T_{g}$. The relaxation time $\tau(T)$ diverges as $T \to T_{g}$ and the algebraic prefactor exponent $x(T)$ decreases from $\approx 0.5$ at high temperatures to $(d-2+\eta)/2z \sim 0.070$ at $T_g$. These equilibrium relaxation results were obtained from high quality simulations on very large samples (up to $L=64$) and have never been improved on since.

Thus the general pattern of the paramagnetic state relaxation reported by Ogielski in the $3d$ ISG and the observed behavior of the SK model described above show striking similarities; above all both sets of data in the paramagnetic region are consistent with stretched exponential decay having an exponent $\beta(T)$ tending to a critical value $\beta_{c} \approx 1/3$ at $T_g$. These two models represent the lowest integer dimension at which finite temperature ISG ordering takes place and the infinite dimension mean field limit (for which the RSB theory is well established) respectively, so it would seem reasonable to expect that the same pattern of relaxation should hold for ISGs at all intermediate dimensions also, both above and below the upper critical dimension (see Ref. \cite{bernardi:94}).

Shortly after Ogielski's work was published it was noted that his data were consistent with $\beta(T_g)=1/3$, and it was conjectured \cite{campbell:85} that stretched exponential relaxation with an exponent tending to precisely $1/3$ at criticality could be universal in ISGs. The argument is briefly summarized in the next section.

\section{Percolation transition scenario}

A thermodynamic phase transition can be regarded as a qualitative change in the topology of the thermodynamically attainable phase space with decreasing temperature, i.e. with decreasing internal energy. Thus for a standard ferromagnetic transition a high temperature spherical phase space becomes more and more "elliptic" as the temperature is lowered and the number of attainable states drops; finally at $T_c$ it splits into two (up and down) mirror-image subspaces. In a na\"{i}ve scenario based on RSB, for an ISG at $T_g$ phase space shatters into a large number of inequivalent clusters. In Euclidean space there is a well studied transition of this type, the percolation transition. For a concentration $p > p_c$ there is a giant cluster of sites while just below $p_c$ there are only small inequivalent clusters.

The total phase space of an $N$-spin $S=1/2$ Ising system is an $N$ dimensional hypercube. Relaxation of any $N$-spin Ising system by successive single spin updates can be considered strictly as a random walk of the system point along near neighbor edges among the thermodynamically attainable vertices on this hypercube  \cite{ogielski:85}.
It was argued \cite{campbell:85} that as random walks on full [hyper]spherical surfaces result in pure exponential decay \cite{debye:29,caillol:04} and random walks on threshold percolation clusters in Euclidean space lead to sub-linear diffusion $\langle R^2 \rangle \sim t^{\beta_{d}}$ \cite{alexander:82}, random walks on threshold
percolation clusters inscribed on [hyper]spheres would be
characterized by "sub-exponential" relaxation of the form
$q(t) = \langle \cos(\theta(t))\rangle =
\exp[-(t/\tau)^{\beta_{d}}]$ with the same exponents $\beta_{d}$ as in
the corresponding Euclidean space. This was demonstrated numerically
for $d = 3$ to $8$ \cite{jund:01}. A hypercube being topologically
equivalent to a hypersphere, on a diluted hypercube
at threshold \cite{eordos:79,borgs:06} random walks can be expected to lead to stretched exponential relaxation with exponent $\beta = 1/3$. Successive explicit numerical studies of random walks on the randomly occupied hypercube at $p_{c}$ \cite{campbell:87,dealmeida:00,lemke:11} have confirmed this.

It is important to note that the limiting behavior for relaxation due to diffusion on sparse graphs has been shown analytically to take the form of a stretched exponential with exponent $\beta = 1/3$ \cite{bray:88,samukhin:08}.

It was further conjectured \cite{campbell:85} that in a complex Ising system such as an ISG the phase transition would be analogous to a percolation transition in configuration space. Detailed "rough landscape" models for the configuration space of complex systems have been widely invoked (see for instance Refs.~\cite{angelani:98,doye:02,burda:07}); these models can be thought of in terms of
linked basins with a gradual dilution of the links leading finally to a percolation threshold. As the critical properties of a percolation transition are very robust, if the configuration space percolation threshold scenario is valid the basic critical behavior should not be sensitive to model details; in particular stretched exponential relaxation with exponent $1/3$ should be observed generically in ISGs whatever the space dimension and possibly also in a wider class of complex systems.

\section{Conclusion}

The present numerical SK results taken together with Ogielski's $3d$ ISG analysis \cite{ogielski:85} can be taken as a strong empirical indication of a universal equilibrium relaxation pattern for ISGs in the paramagnetic regime : stretched exponential decay Eq.~\ref{qt} having an exponent $\beta(T)$ which tends to $1/3$ when the relaxation time diverges at the critical temperature $T_{g}$. Auto-correlation function decay at the percolation transition on the randomly occupied hypercube \cite{campbell:87,dealmeida:00,lemke:11} has precisely this form, suggesting that the ISG phase transition can indeed be considered in terms of a percolation transition in phase space.

The same type of relaxation pattern has been observed in many other complex system studies, both numerical and experimental. It is tempting to conclude that the physical scenario could have implications for a wide class of transitions, extending well beyond the ISG family.

\section{Acknowledgement} We would like to thank Ney Lemke for an instructive discussion and for communicating his unpublished data.


\begin{thebibliography}{99}
\bibitem{beta} We follow well established conventions where unfortunately $\beta$ is traditionally used to represent a number of quite different parameters : the inverse of the temperature $1/T$, a critical exponent, and following \cite{ogielski:85}, a stretched exponential exponent. We hope that it will be clear from context which parameter is referred to.
\end{thebibliography}
\end{document}